\documentclass[twocolumn]{aastex631}

\usepackage{amsmath}
\usepackage{amssymb}

\begin{document}

\title{ODIN: Spectroscopic Validation of Ly$\alpha$-Emitting Galaxy Samples with DESI}

\author[0009-0003-5244-3700]{Ethan Pinarski}
\affiliation{Department of Physics and Astronomy, Purdue University, 525 Northwestern Avenue, West Lafayette, IN 47907, USA}

\author[0009-0008-1418-6981]{Govind Ramgopal}
\affiliation{Physics and Astronomy Department, Rutgers, The State University, Piscataway, NJ 08854, USA}

\author[0000-0002-9811-2443]{Nicole Firestone}
\affiliation{Physics and Astronomy Department, Rutgers, The State University, Piscataway, NJ 08854, USA}

\author[0000-0003-3004-9596]{Kyoung-Soo Lee}
\affiliation{Department of Physics and Astronomy, Purdue University, 525 Northwestern Avenue, West Lafayette, IN 47907, USA}

\author[0000-0003-1530-8713]{Eric Gawiser}
\affiliation{Physics and Astronomy Department, Rutgers, The State University, Piscataway, NJ 08854, USA}

\author[0000-0002-4928-4003]{Arjun Dey}
\affiliation{NSF's National Optical-Infrared Astronomy Research Laboratory, 950 N. Cherry Ave., Tucson, AZ 85719, USA}

\author[0000-0001-5999-7923]{A.~Raichoor}
\affiliation{Lawrence Berkeley National Laboratory, 1 Cyclotron Road, Berkeley, CA 94720, USA}

\author[0000-0001-5567-1301]{Francisco Valdes}
\affiliation{NSF's National Optical-Infrared Astronomy Research Laboratory, 950 N. Cherry Ave., Tucson, AZ 85719, USA}

\author[0000-0002-1328-0211]{Robin Ciardullo}
\affiliation{Department of Astronomy \& Astrophysics, The Pennsylvania State University, University Park, PA 16802, USA}
\affiliation{Institute for Gravitation and the Cosmos, The Pennsylvania State University, University Park, PA 16802, USA}

\author[0000-0003-0822-452X]{Jessica N. ~Aguilar}
\affiliation{Lawrence Berkeley National Laboratory, 1 Cyclotron Road, Berkeley, CA 94720, USA}

\author[0000-0001-6098-7247]{S.~Ahlen}
\affiliation{Department of Physics, Boston University, 590 Commonwealth Avenue, Boston, MA 02215 USA}

\author[0000-0001-9712-0006]{D.~Bianchi}
\affiliation{Dipartimento di Fisica ``Aldo Pontremoli'', Universit\`a degli Studi di Milano, Via Celoria 16, I-20133 Milano, Italy}
\affiliation{INAF-Osservatorio Astronomico di Brera, Via Brera 28, 20122 Milano, Italy}

\author{D.~Brooks}
\affiliation{Department of Physics \& Astronomy, University College London, Gower Street, London, WC1E 6BT, UK}

\author[0000-0001-7316-4573]{F.~J.~Castander}
\affiliation{Institut d'Estudis Espacials de Catalunya (IEEC), c/ Esteve Terradas 1, Edifici RDIT, Campus PMT-UPC, 08860 Castelldefels, Spain}
\affiliation{Institute of Space Sciences, ICE-CSIC, Campus UAB, Carrer de Can Magrans s/n, 08913 Bellaterra, Barcelona, Spain}

\author[0009-0000-9347-1933]{M. Candela Cerdosino}
\affiliation{Instituto de Astronomía Teórica y Experimental (IATE), CONICET-UNC, Laprida 854, X500BGR, Córdoba, Argentina}
\affiliation{Facultad de Matemática, Astronomía, Física y Computación, Universidad Nacional de Córdoba, Bvd. Medina Allende s/n, Ciudad Universitaria, X5000HU, Córdoba, Argentina}

\author{T.~Claybaugh}
\affiliation{Lawrence Berkeley National Laboratory, 1 Cyclotron Road, Berkeley, CA 94720, USA}

\author[0000-0002-2169-0595]{A.~Cuceu}
\affiliation{Lawrence Berkeley National Laboratory, 1 Cyclotron Road, Berkeley, CA 94720, USA}

\author[0000-0002-0553-3805]{K.~S.~Dawson}
\affiliation{Department of Physics and Astronomy, The University of Utah, 115 South 1400 East, Salt Lake City, UT 84112, USA}

\author[0000-0002-1769-1640]{A.~de la Macorra}
\affiliation{Instituto de F\'{\i}sica, Universidad Nacional Aut\'{o}noma de M\'{e}xico,  Circuito de la Investigaci\'{o}n Cient\'{\i}fica, Ciudad Universitaria, Cd. de M\'{e}xico  C.~P.~04510,  M\'{e}xico}

\author{P.~Doel}
\affiliation{Department of Physics \& Astronomy, University College London, Gower Street, London, WC1E 6BT, UK}

\author[0000-0003-4992-7854]{S.~Ferraro}
\affiliation{Lawrence Berkeley National Laboratory, 1 Cyclotron Road, Berkeley, CA 94720, USA}
\affiliation{University of California, Berkeley, 110 Sproul Hall \#5800 Berkeley, CA 94720, USA}

\author[0000-0002-3033-7312]{A.~Font-Ribera}
\affiliation{Instituci\'{o} Catalana de Recerca i Estudis Avan\c{c}ats, Passeig de Llu\'{\i}s Companys, 23, 08010 Barcelona, Spain}
\affiliation{Institut de F\'{i}sica d'Altes Energies (IFAE), The Barcelona Institute of Science and Technology, Edifici Cn, Campus UAB, 08193, Bellaterra (Barcelona), Spain}

\author[0000-0002-2890-3725]{J.~E.~Forero-Romero}
\affiliation{Departamento de F\'isica, Universidad de los Andes, Cra. 1 No. 18A-10, Edificio Ip, CP 111711, Bogot\'a, Colombia}
\affiliation{Observatorio Astron\'omico, Universidad de los Andes, Cra. 1 No. 18A-10, Edificio H, CP 111711 Bogot\'a, Colombia}

\author[0000-0001-9632-0815]{E.~Gaztañaga}
\affiliation{Institut d'Estudis Espacials de Catalunya (IEEC), c/ Esteve Terradas 1, Edifici RDIT, Campus PMT-UPC, 08860 Castelldefels, Spain}
\affiliation{Institute of Cosmology and Gravitation, University of Portsmouth, Dennis Sciama Building, Portsmouth, PO1 3FX, UK}
\affiliation{Institute of Space Sciences, ICE-CSIC, Campus UAB, Carrer de Can Magrans s/n, 08913 Bellaterra, Barcelona, Spain}

\author[0000-0003-3142-233X]{S.~Gontcho A Gontcho}
\affiliation{University of Virginia, Department of Astronomy, Charlottesville, VA 22904, USA}

\author[0000-0002-4902-0075]{Lucia Guaita}
\affiliation{Departamento de Ciencias Fisicas, Universidad Andres Bello, Fernandez Concha 700, Las Condes, Santiago, Chile}

\author[0000-0003-0825-0517]{G.~Gutierrez}
\affiliation{Fermi National Accelerator Laboratory, PO Box 500, Batavia, IL 60510, USA}

\author[0000-0001-8221-8406]{Stephen Gwyn}
\affiliation{Herzberg Astronomy and Astrophysics Research Centre, National Research Council of Canada, Victoria, British Columbia, Canada}

\author[0000-0002-9136-9609]{H.~K.~Herrera-Alcantar}
\affiliation{Institut d'Astrophysique de Paris. 98 bis boulevard Arago. 75014 Paris, France}
\affiliation{IRFU, CEA, Universit\'{e} Paris-Saclay, F-91191 Gif-sur-Yvette, France}

\author[0000-0003-3428-7612]{Ho Seong Hwang}
\affiliation{Astronomy Program, Department of Physics and Astronomy, Seoul National University, 1 Gwanak-ro, Gwanak-gu, Seoul 08826, Republic of Korea} 
\affiliation{SNU Astronomy Research Center, Seoul National University, 1 Gwanak-ro, Gwanak-gu, Seoul 08826, Republic of Korea} 

\author[0000-0003-0201-5241]{R.~Joyce}
\affiliation{NSF NOIRLab, 950 N. Cherry Ave., Tucson, AZ 85719, USA}

\author[0000-0002-0000-2394]{S.~Juneau}
\affiliation{NSF NOIRLab, 950 N. Cherry Ave., Tucson, AZ 85719, USA}

\author{R.~Kehoe}
\affiliation{Department of Physics, Southern Methodist University, 3215 Daniel Avenue, Dallas, TX 75275, USA}

\author[0000-0002-8828-5463]{D.~Kirkby}
\affiliation{Department of Physics and Astronomy, University of California, Irvine, 92697, USA}

\author[0000-0003-3510-7134]{T.~Kisner}
\affiliation{Lawrence Berkeley National Laboratory, 1 Cyclotron Road, Berkeley, CA 94720, USA}

\author[0000-0001-6356-7424]{A.~Kremin}
\affiliation{Lawrence Berkeley National Laboratory, 1 Cyclotron Road, Berkeley, CA 94720, USA}

\author[0000-0001-6270-3527]{Ankit Kumar}
\affiliation{Universidad Andres Bello, Facultad de Ciencias Exactas, Departamento de Fisica y Astronomia, Instituto de Astrofisica, Fernandez Concha 700, Las Condes, Santiago RM, Chile}

\author[0000-0002-6731-9329]{C.~Lamman}
\affiliation{The Ohio State University, Columbus, 43210 OH, USA}

\author[0000-0003-1838-8528]{M.~Landriau}
\affiliation{Lawrence Berkeley National Laboratory, 1 Cyclotron Road, Berkeley, CA 94720, USA}

\author[0000-0001-7178-8868]{L.~Le~Guillou}
\affiliation{Sorbonne Universit\'{e}, CNRS/IN2P3, Laboratoire de Physique Nucl\'{e}aire et de Hautes Energies (LPNHE), FR-75005 Paris, France}

\author[0000-0003-1887-1018]{M.~E.~Levi}
\affiliation{Lawrence Berkeley National Laboratory, 1 Cyclotron Road, Berkeley, CA 94720, USA}

\author[0000-0002-4623-0683]{Yufeng Luo}
\affiliation{$^1$Department of Physics and Astronomy,
University of Wyoming,
1000 E. University Ave., Laramie, WY 82071, USA}

\author[0000-0003-4962-8934]{M.~Manera}
\affiliation{Departament de F\'{i}sica, Serra H\'{u}nter, Universitat Aut\`{o}noma de Barcelona, 08193 Bellaterra (Barcelona), Spain}
\affiliation{Institut de F\'{i}sica d’Altes Energies (IFAE), The Barcelona Institute of Science and Technology, Edifici Cn, Campus UAB, 08193, Bellaterra (Barcelona), Spain}

\author[0000-0002-4279-4182]{P.~Martini}
\affiliation{Center for Cosmology and AstroParticle Physics, The Ohio State University, 191 West Woodruff Avenue, Columbus, OH 43210, USA}
\affiliation{Department of Astronomy, The Ohio State University, 4055 McPherson Laboratory, 140 W 18th Avenue, Columbus, OH 43210, USA}

\author[0000-0002-1125-7384]{A.~Meisner}
\affiliation{NSF NOIRLab, 950 N. Cherry Ave., Tucson, AZ 85719, USA}

\author[0000-0002-6610-4836]{R.~Miquel}
\affiliation{Instituci\'{o} Catalana de Recerca i Estudis Avan\c{c}ats, Passeig de Llu\'{\i}s Companys, 23, 08010 Barcelona, Spain}
\affiliation{Institut de F\'{i}sica d’Altes Energies (IFAE), The Barcelona Institute of Science and Technology, Edifici Cn, Campus UAB, 08193, Bellaterra (Barcelona), Spain}

\author[0000-0002-2733-4559]{J.~Moustakas}
\affiliation{Department of Physics and Astronomy, Siena University, 515 Loudon Road, Loudonville, NY 12211, USA}

\author{A.~D.~Myers}
\affiliation{Department of Physics \& Astronomy, University  of Wyoming, 1000 E. University, Dept.~3905, Laramie, WY 82071, USA}

\author[0000-0001-9070-3102]{S.~Nadathur}
\affiliation{Institute of Cosmology and Gravitation, University of Portsmouth, Dennis Sciama Building, Portsmouth, PO1 3FX, UK}

\author[0000-0002-0905-342X]{Gautam R. Nagaraj}
\affiliation{Department of Astronomy \& Astrophysics, The Pennsylvania State University, University Park, PA 16802, USA}
\affiliation{Institute for Gravitation and the Cosmos, The Pennsylvania State University, University Park, PA 16802, USA}

\author[0000-0003-3188-784X]{N.~Palanque-Delabrouille}
\affiliation{IRFU, CEA, Universit\'{e} Paris-Saclay, F-91191 Gif-sur-Yvette, France}
\affiliation{Lawrence Berkeley National Laboratory, 1 Cyclotron Road, Berkeley, CA 94720, USA}

\author[0000-0001-9521-6397]{Changbom Park}
\affiliation{Korea Institute for Advanced Study, 85 Hoegi-ro, Dongdaemun-gu, Seoul 02455, Republic of Korea}

\author[0000-0002-0644-5727]{W.~J.~Percival}
\affiliation{Department of Physics and Astronomy, University of Waterloo, 200 University Ave W, Waterloo, ON N2L 3G1, Canada}
\affiliation{Perimeter Institute for Theoretical Physics, 31 Caroline St. North, Waterloo, ON N2L 2Y5, Canada}
\affiliation{Waterloo Centre for Astrophysics, University of Waterloo, 200 University Ave W, Waterloo, ON N2L 3G1, Canada}

\author[0000-0001-6979-0125]{I.~P\'erez-R\`afols}
\affiliation{Departament de F\'isica, EEBE, Universitat Polit\`ecnica de Catalunya, c/Eduard Maristany 10, 08930 Barcelona, Spain}

\author[0000-0001-7145-8674]{F.~Prada}
\affiliation{Instituto de Astrof\'{i}sica de Andaluc\'{i}a (CSIC), Glorieta de la Astronom\'{i}a, s/n, E-18008 Granada, Spain}

\author{G.~Rossi}
\affiliation{Department of Physics and Astronomy, Sejong University, 209 Neungdong-ro, Gwangjin-gu, Seoul 05006, Republic of Korea}

\author[0000-0002-9646-8198]{E.~Sanchez}
\affiliation{CIEMAT, Avenida Complutense 40, E-28040 Madrid, Spain}

\author[0000-0002-7712-7857]{Marcin Sawicki}
\affiliation{Institute for Computational Astrophysics and Department of Astronomy and Physics, Saint Mary's University, 923 Robie St., Halifax, Nova Scotia, B3H 3C3,
Canada}

\author{D.~Schlegel}
\affiliation{Lawrence Berkeley National Laboratory, 1 Cyclotron Road, Berkeley, CA 94720, USA}

\author{M.~Schubnell}
\affiliation{Department of Physics, University of Michigan, 450 Church Street, Ann Arbor, MI 48109, USA}
\affiliation{University of Michigan, 500 S. State Street, Ann Arbor, MI 48109, USA}

\author[0000-0002-3461-0320]{J.~Silber}
\affiliation{Lawrence Berkeley National Laboratory, 1 Cyclotron Road, Berkeley, CA 94720, USA}

\author[0000-0002-4362-4070]{Hyunmi Song}
\affiliation{Department of Astronomy and Space Science, Chungnam National University, 99 Daehak-ro, Yuseong-gu, Daejeon 34134, Republic of Korea}

\author{D.~Sprayberry}
\affiliation{NSF NOIRLab, 950 N. Cherry Ave., Tucson, AZ 85719, USA}

\author[0000-0003-1704-0781]{G.~Tarl\'{e}}
\affiliation{University of Michigan, 500 S. State Street, Ann Arbor, MI 48109, USA}

\author[0000-0001-6162-3023]{Paulina Troncoso Iribarren}
\affiliation{Facultad de Ingeniería y Arquitectura, Universidad Central de Chile, Avenida Francisco de Aguirre 0405, 171-0614 La Serena, Coquimbo, Chile}

\author{B.~A.~Weaver}
\affiliation{NSF NOIRLab, 950 N. Cherry Ave., Tucson, AZ 85719, USA}

\author[0000-0003-3078-2763]{Yujin Yang}
\affiliation{Korea Astronomy and Space Science Institute, 776 Daedeokdae-ro, Yuseong-gu, Daejeon 34055, Republic of Korea}

\author[0000-0001-6047-8469]{Ann Zabludoff}
\affiliation{Steward Observatory, University of Arizona, 933 North Cherry Avenue, Tucson, AZ 85721, USA}





\begin{abstract}
The One-hundred-deg$^2$ DECam Imaging in Narrowbands (ODIN) survey is conducting the widest-field deep narrow-band imaging of the equatorial and southern skies. ODIN uses three custom-built narrow-band (NB) filters that sample Ly$\alpha$-emitting galaxies (LAEs) within thin cosmic slices centered at $z=2.4$, 3.1, and 4.5. In this work, we utilize extensive DESI spectroscopy of ODIN-selected galaxies in the COSMOS and XMM-LSS fields to validate our LAE selection.  2--4~hr exposures with DESI yielded redshift confirmation of 3,075 ODIN LAE candidates with NB magnitudes brighter than 26~mag. Restricting to objects that yield high-confidence redshifts, the confirmation rates are (93, 96, 92)\% at $z=$(2.4, 3.1, 4.5). The primary contaminants consist of active galactic nuclei at the expected Ly$\alpha$ redshift range and lower redshifts (C~{\sc iv}, C~{\sc iii}]), with the remainder being star-forming galaxies ([O~{\sc ii}] and  [O~{\sc iii}]). We find minimal contamination from [O~{\sc ii}] emitters in our sample ($\lesssim$1\%), implying that our ${\rm REW}>20$~\AA\ narrow-band excess photometry requirement is sufficient to remove them. 
\end{abstract}

\section{Introduction}\label{sec:intro}

Emission lines in galaxies offer rich diagnostics of the physical state of the gas -- revealing its density, temperature, kinematics, and chemical composition -- as well as key properties of the host galaxies, such as star formation rates (SFR), star formation histories, and dust content \textbf{\citep{kewley2019}}. They also serve as powerful tools for identifying active galactic nuclei (AGN) and probing their characteristics, including accretion rates and central blackhole masses. Using these diagnostics, wide-field spectroscopic surveys have played a crucial role in tracing the evolution of galaxy and AGN activity across cosmic time \citep[e.g.,][]{reddy18,sanders21}. Existing studies show that galaxies exhibiting strong emission lines tend to have low stellar mass, high specific SFR, low dust content, and low gas metallicity \citep[e.g.,][]{cardamone09,indahl21}, suggesting that they are among the least biased tracers of the underlying matter distribution. With the advent of wide-field spectrographs, emission-line galaxies are promising targets to study the large-scale structure and expansion history of the universe \citep[e.g.,][]{drinkwater2010,gebhardt21,raichoor23}.

Lyman-alpha (Ly$\alpha$) emission at a rest-frame wavelength of 1215.67~\AA\ has proven especially valuable for exploring the $z > 2$ universe. As the hydrogen recombination line corresponding to the $n=2 \rightarrow 1$ transition, it is inherently the most luminous line emitted from star-forming regions in galaxies. Its distinctively asymmetric line profile aids in making it readily distinguishable from other emission lines, provided the signal-to-noise ratio (S/N) is sufficiently high \citep{partridge&peebles1967, ouchi2020, hayes2015}. At $z \gtrsim 2$, Ly$\alpha$ is redshifted into the optical window, enabling ground-based telescopes to efficiently detect Ly$\alpha$-emitting galaxies (LAEs) and use them to trace large-scale structure. 

Indeed, surveys such as the Hobby-Eberly Telescope Dark Energy Experiment \citep[HETDEX:][]{gebhardt21} leverage LAEs to constrain the properties of dark energy out to $z \approx 3$. The second stage of the Dark Energy Spectroscopic Instrument {\citep[DESI-2:][]{schlegel2022}} will follow suit. Moreover, the spatial distribution of LAEs provides a statistical connection between galaxies and their dark matter halos, enabling studies of galaxy evolution across cosmic time \citep[e.g.,][]{white2024clustering,herrera25}. LAEs also serve as signposts of massive cosmic structures—overdensities that will eventually evolve into virialized galaxy clusters—long before these structures become fully formed \citep[e.g.,][]{lee24,ramakrishnan24}.

For the past two decades, wide-field imagers have enabled highly efficient photometric selection of LAEs by identifying sources with a significant narrow-band flux excess relative to the broad-band continuum \citep[e.g.,][]{Gronwall_2007ApJ, Ouchi_2008ApJS, Zheng_2014MNRAS}. This technique has proven effective in assembling large, uniformly selected samples of LAEs for both astrophysical and cosmological studies \citep[e.g.,][]{kikuta2023silverrush, firestone24, Zhang_2025ApJ}. 

The One-hundred-deg$^2$ DECam Imaging in Narrowbands (ODIN; \citealt{lee24}) is the largest photometric LAE survey to date. ODIN is a NOIRLab program conducted with the Dark Energy Camera (DECam) on the Blanco 4m Telescope at Cerro Tololo, employing custom narrowband filters (the full-width-at-half-maximum of $\sim$ 75--100~\AA) designed to isolate Ly$\alpha$ emission at three evenly spaced epochs during Cosmic Noon: $N419$ ($z\approx2.4$), $N501$ ($z\approx3.1$), and $N673$ ($z\approx4.5$). Upon completion, ODIN will cover up to 91~deg$^2$ in each filter across seven deep, wide fields to a depth of $\sim$25.7~AB, yielding an expected sample of $\sim$100,000 LAEs.

The inaugural ODIN LAE sample in the extended COSMOS field was presented in \citet{firestone24}, along with refined selection techniques aimed at improving sample purity and reducing contamination from interlopers. ODIN's LAE catalog has since enabled studies on a broad range of topics, including the robust selection of protoclusters \citep{ramakrishnan24, ramakrishnan25a, ramakrishnan25b}, the physical association of protoclusters and Ly$\alpha$ blobs \citep{ramakrishnan23}, star formation in LAEs \citep{firestone25, cerdosino25}, the Ly$\alpha$ luminosity function \citep{nagaraj25}, clustering properties of LAEs and blobs \citep{herrera25,Moon25b}, and Ly$\alpha$ emission-line profiles \citep{uzsoy25b}. A critical remaining step is to accurately quantify the contamination rate from interlopers—such as lower-redshift star-forming galaxies or AGN—that can masquerade as LAEs and compromise sample purity. Such a quantitative assessment is essential both to validate these results and to inform future emission-line galaxy surveys.

In this paper, we present a systematic assessment of the ODIN LAE sample based on DESI spectroscopic follow-up observations in the COSMOS and XMM-LSS fields (details from DESI provided in A. Dey et al. in prep). These observations -- which we refer to as DESI-ODIN observations, hereafter -- were motivated to better prepare for DESI-2, which will target $z \gtrsim 2$ LAEs as cosmological tracers. These data have been used to measure the LAE properties, such as clustering \citep{white2024clustering}, the dependence of the Ly$\alpha$ profile on the local environment \citep{uzsoy25}, and the development of an automated spectroscopic classifier and redshift fitter \citep{uzsoy25b}.

This paper is organized as follows. In Section~\ref{sec:data_selection}, we provide the details of target selection and DESI observations. Section~\ref{sec:results} focuses on the contamination rate, completeness, and demographics of ODIN-selected LAEs, followed by conclusions in Section~\ref{sec:conclusions}. Throughout this paper, we adopt a cosmology with
$\Omega_m=0.27$, $\Omega_\Lambda=0.73$, $H_0 = 70$~km~s$^{-1}$~Mpc$^{-1}$. All magnitudes are given in the AB system \citep{oke83}.

\section{Target Selection and Follow-up Spectroscopy}\label{sec:data_selection}

Our primary goal is to quantify the robustness of the ODIN LAE samples. Although the DESI-ODIN observations present a unique opportunity to directly estimate the purity of our LAE samples, it is important to note key differences in the photometric properties of the ODIN LAEs and the full set of targets of the DESI-ODIN observations. The latter will be presented in A. Dey et al. (in prep). While both selections rely on the ODIN narrow-band (NB) imaging data to isolate Ly$\alpha$ emission falling into the $N419$, $N501$, and $N673$ bands to select LAEs at $z=2.4$, 3.1, and 4.5, respectively, the programs diverge in the ways source detection and photometry are performed, the broad-band imaging dataset used in the selection, and the strictness of the narrow-band excess flux density requirement that sets a minimum value for the emission line equivalent width. Here, we summarize both selections, emphasizing the key distinctions.

\subsection{ODIN LAE Selection}\label{subsec:odin-lae}

The full detail of the ODIN LAE selection is given in \citet{firestone24}. Briefly, it uses photometric catalogs constructed using the Source Extractor (SE) software \citep{SE}, where one NB image is used as the detection band to detect the source while $ugrizy$ and other NB images are used as measurement bands to collect additional photometric data. Only HSC-SSP and the corresponding CLAUDS $u$-band data are used \citep{sawicki19}. Photometry is measured in 2" diameter circular apertures and then corrected by an appropriate factor determined based on the empirical point spread function created from bright unsaturated stars in the field, with an increased estimate of total photometry applied for clearly extended sources.  

Unlike the DESI-ODIN LAE selection, ODIN uses a double-broad-band continuum estimate in conjunction with a given NB to compute the colors; this is designed to accurately estimate the NB excess by taking into account the continuum slope \citep{firestone24}. The color criteria used for all three redshift bins correspond to a minimum rest-frame equivalent width of 20~\AA. Finally, HSC-SSP-generated star masks \citep{coupon18} are applied to the LAE sample to minimize image artifacts near bright stars being classified as LAEs. Doing so removes approximately 20\% of the candidates. 

\begin{figure*}
    \centering
    \includegraphics[width=0.95\linewidth, 
    ]{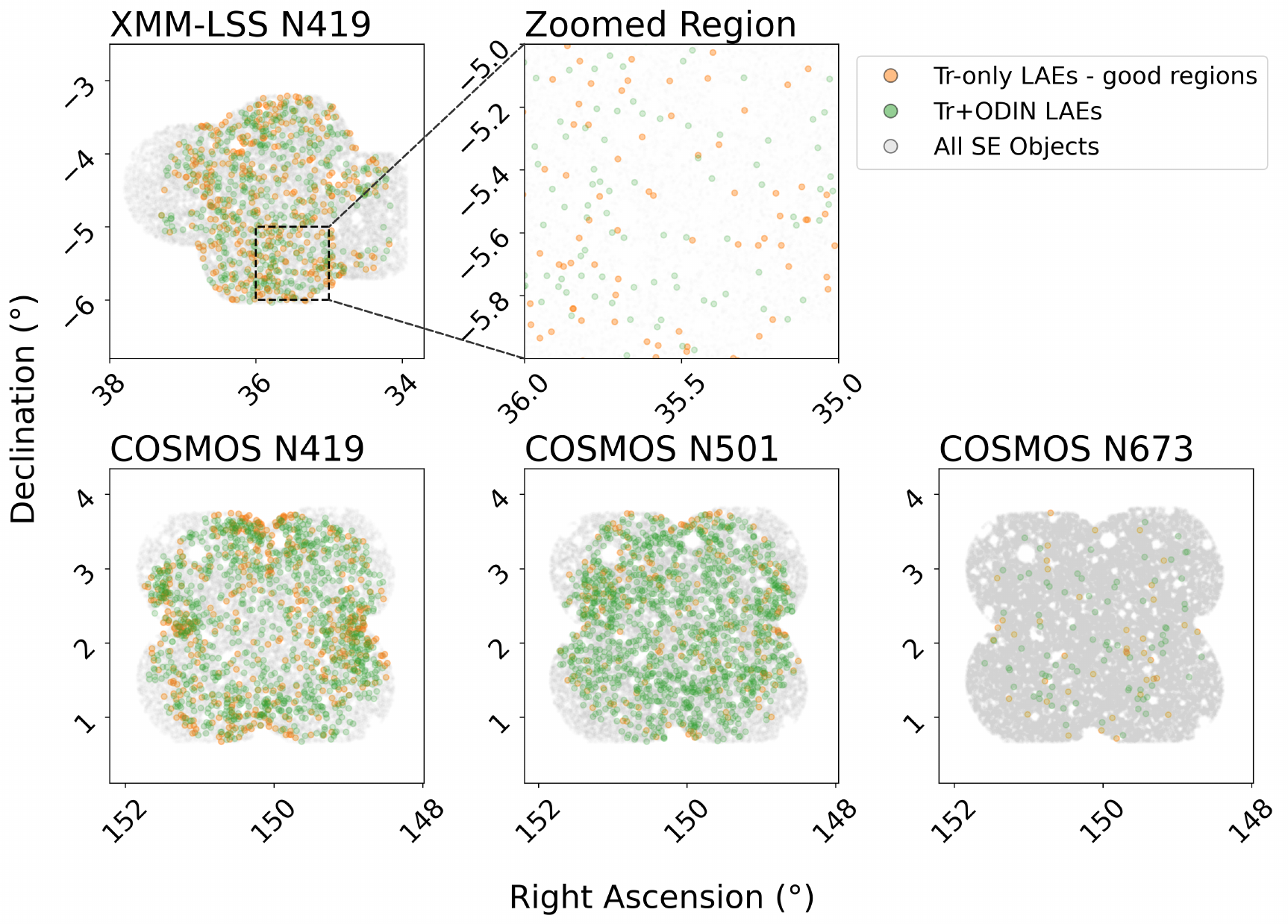}
    \caption{Maps of two ODIN fields (COSMOS and XMM-LSS) showing the locations of spectroscopically confirmed LAEs. The objects are colored as follows: All SE objects in good regions (gray); Tr-only LAEs in good regions (not starmasked and in regions containing high-quality HSC Broadband data) (orange); and Tr+ODIN LAEs (green). Of the confirmed LAEs not selected by ODIN’s criteria, roughly half lie in sky regions excluded by ODIN (either in starmasks or regions without HSC broad-band data). Counts of the subsets of Tr-only LAEs can be found in \autoref{tab:spec-lae-distribution}.}
    \label{fig:fig1}
\end{figure*}

\subsection{DESI-ODIN LAE Selection}\label{subsec:desi-odin-lae}

Dey et al. (in prep.) give a comprehensive description of the DESI-ODIN LAE selection, which served as the parent sample for their targets. The DESI-ODIN target selection makes use of $grz$ imaging from the DESI Imaging Legacy Surveys \citep[][]{dey19} together with the much deeper $grizy$ data from the Hyper-Suprime Cam Subaru Strategic Program \citep[HSC-SSP, hereafter:][]{ssp_dr1} in the COSMOS and XMM-LSS fields. While this approach compromises uniformity in photometric depth across the field, it enables target selection over the full DESI field of view, which is larger than the region covered by HSC-SSP. In comparison, the NB imaging depth provided by the ODIN survey is uniform within 5\% \citep{lee24}.

The DESI-ODIN photometric catalogs are constructed using the {\tt Tractor} software \citep{lang16}, where a given NB image is used as the detection image while photometric measurements are made on images taken with the remaining filters. As {\tt Tractor} performs forced model-based photometry, all measured fluxes are {\it total}, requiring no aperture correction. Similar to the conventional LAE selection in the literature \citep[e.g.,][]{Hu_McMahon_1996Natur, gawiser2006physical, zheng2017first, kikuta2023silverrush},  color and magnitude cuts are imposed to isolate a strong line emission falling into the designated NB filter at a S/N of 5 or greater. Star masking is also applied on the DESI-ODIN sample.

The narrow-to-broad-band color criteria adopted by DESI for the candidate selection tend to be more inclusive than ODIN, in the sense that the minimum color excess and thus the minimum rest-frame equivalent width is lower. This decision was driven by the desire to understand how the success rate varies as a function of colors. As a result, the total number of sources targeted as potential LAEs by DESI at each redshift is significantly larger than that selected by ODIN, which allows us to estimate the incompleteness of the ODIN LAE selection. Hereafter, we refer to the former as Tr LAEs, to denote their selection from the {\tt Tractor} catalog, and the latter as ODIN LAEs. Direct comparisons of the two samples are presented in Section~\ref{subsec:comparison}.

\subsection{DESI Observations and Visual Inspection}\label{subsec:desi_vi}

DESI is a wide-field, fiber-fed spectrograph mounted on the Mayall 4 m telescope at Kitt Peak National Observatory \citep{DESI2022:instrumentation}. With the capability of obtaining spectra of up to $\sim$5000 sources per pointing \citep{DESI2016:instrument_design,DESI2024:optical_corrector,DESI2024:fiber_system}, DESI is currently carrying out a spectroscopic survey over $\sim$17,000~deg$^2$ of sky \citep{DESI2023:survey,DESI2025:DR1} with the primary aim of constraining the nature of dark energy \citep{DESI2025:cosmology,DESI2025:BAO}.

Details of the DESI follow-up observations of the Tr LAE sample will be presented in A. Dey et al. (in prep.); here we summarize the key aspects relevant to this work. DESI spectroscopy was obtained in three observing campaigns conducted in March 2022, December 2022–January 2023, and April 2023. A subset of the Tr LAEs (see Section~\ref{subsec:desi-odin-lae}) was targeted across these campaigns with varying effective exposure times.

During the March 2022 run, which focused on $N501$ and $N673$ LAEs in COSMOS, these sources were assigned the highest priority and observed as primary targets, achieving a median effective exposure time of 2.75~hr. The second campaign, targeting $N419$ LAEs in XMM-LSS, divided sources into high- and low-priority classes, with the high-priority subset receiving approximately twice the exposure time (effective time of 3.2~hr) of the low-priority subset. The final campaign in April 2023 observed a mixture of samples, not only including ODIN $N419$, $N501$, and $N673$ LAEs but also those selected based on other datasets. Although $\approx$85\% of these targets obtained effective exposure times of at least $1.3$~hr, the observations of the remaining 15\% were shallower, resulting in lower S/N values. We examine the impact of including or excluding these lower-S/N sources on the inferred sample purity in Section~\ref{subsec:purity}. All spectra were reduced using the standard DESI pipeline \citep{DESI2023:specz_pipeline}.

To determine an accurate redshift and spectral type for the DESI-ODIN targets, a visual classification campaign was conducted from June 2023 to May 2024. The comprehensive details are presented in A. Dey et al. (in prep), and we provide a brief overview here.

In the visual classification campaign, two or more experts were assigned to each of the DESI-ODIN targets' spectrum. They evaluated the redshift, spectrum quality ($\tt q$), and spectral type based on prevalent emission line features in each spectrum. $\tt q$ refers to the confidence in the accuracy of a spectrum's assessed redshift and spectral type given as an integer from 0-4 and is determined by the number and S/N of emission line features as follows: $\tt q = 0,1$: spectrum has no apparent features, $\tt q = 2$: possible feature(s) but not certain, $\tt q = 3$: one prominent feature present, $\tt q = 4$: two or more features present.
The final $\tt q$ for each target is a weighted average of the individual expert assessments.

In this work, we will define DESI-ODIN targets with ${\tt q} \geq 2.5$ as spectroscopically confirmed. The cut-off is at ${\tt q} \geq 2.5$ and not ${\tt q} \geq 3$ due to the spectra of LAEs only having the one Ly$\alpha$ emission line feature, so the highest quality rating that could be assigned is ${\tt q} = 3$. A cut-off of ${\tt q} \geq 2.5$ means that at least half of the evaluations assigned to a given target were confident in the quality of their assessment. 

We define DESI-ODIN targets as the subset of Tr LAEs that were placed in fibers.  We further define spec-LAEs as the subset of DESI-ODIN targets that yielded a high-confidence redshift in the expected redshift range and were not found to be AGN; these are the spectroscopically confirmed LAEs shown in \autoref{fig:fig1} and summarized in \autoref{tab:summary}. The number of spectroscopically confirmed targets changes little between ${\tt q} \geq 2.5$ and ${\tt q} \geq 3$.

\subsection{Differences Between the LAE Samples}\label{subsec:comparison}

Based on the selection approaches for Tr- and ODIN LAEs described above, we define three disjoint categories of LAEs:  
\begin{enumerate}
    \item \textbf{Tr-only LAEs:}  objects selected as Tr LAEs based on their photometry in the Tractor catalog but not selected as ODIN LAEs.  
    \item \textbf{Tr+ODIN LAEs:}  objects selected as LAEs based on both their Tractor and SE photometry.  
    \item \textbf{ODIN-only LAEs:}  objects selected as ODIN LAEs based on their photometry in the SE catalog but not selected as Tr LAEs.  
\end{enumerate}

Based on these definitions, the Tr LAEs represent the union of categories 1 and 2, and the ODIN LAEs represent the union of categories 2 and 3.    

To evaluate the effectiveness of ODIN’s LAE selection criteria, we first need to exclude Tr-only LAEs that ODIN could not have selected. These objects reside in regions with HSC starmasks or in regions with no available HSC broad-band data. This leaves behind a subset of \textit{Tr-only LAEs - good regions} for comparison. These objects were not selected as ODIN LAEs due to the narrowband excess threshold criteria (\S~2.2), as we will later explore. A breakdown of the subsets of Tr-only LAEs is presented in \autoref{tab:spec-lae-distribution}. Here, we can see that roughly half of the DESI-identified LAEs that ODIN did not select lie in regions excluded a priori by ODIN.

In Figure~\ref{fig:fig1}, we illustrate the differences between the samples by showing the spatial distribution of Tr and ODIN LAEs. The orange points represent Tr-only LAEs that are in regions with high quality HSC broadband data and not starmasked (\textit{good regions}), and the green points represent Tr+ODIN LAEs. The grey points represent the locations of 10\% of all NB-detected sources within the HSC-SSP broad-band footprint after the application of starmasks shown as white circular regions devoid of sources \citep{coupon18}. As can be seen in the zoomed-in view of the XMM-LSS field (the top right panel of Figure~\ref{fig:fig1}), there are smaller star masks not visible in the full view. In addition, there is a region with no DESI follow-up in the XMM-LSS field at $34.25^\circ<\alpha<39.25^\circ$, $-5.3^\circ<\delta<-4.6^\circ$ due to a difference in survey targeting priorities.

The spatial distribution of the ODIN and Tr LAE targets is indicative of the two projects' respective survey geometries. As described in Sections~\ref{subsec:odin-lae} and \ref{subsec:desi-odin-lae}, ODIN uses HSC-SSP broad-band data, which results in four-leaf-clover-shaped LAE distributions. In contrast, DESI selects objects in a 1.9$^\circ$ radius around each field's center and uses a combination of its Legacy Survey data and HSC imaging for broad-band coverage, resulting in a circular distribution of spectroscopic targets. 

\begin{table*}[ht!]
\begin{center}
    \begin{tabular}{lccccc}
    \hline
     & XMM-LSS $N419$ & COSMOS $N419$ & COSMOS $N501$ & COSMOS $N673$ & total \\
    \hline\hline
     Redshift  & $2.45\pm0.03$ & $2.45\pm0.03$ & $3.12\pm0.03$ & $4.55\pm0.04$  & -\\
     \# of DESI LAEs targeted          & 4,296         & 3,573 & 2,939 & 791  & 11,599 \\
     \# of DESI targets with (${\tt q}\geq 2.5$)$^*$   & 2,983  & 2,385 & 2,358 & 247  & 7,973  \\
     \hline
     \multicolumn{6}{c}{ODIN LAEs selected by \citet{firestone24} } \\
     \hline     
     \# of ODIN LAEs targeted                 & 678      & 1,311 & 1,422 & 94 & 3,505 \\
     \# of ODIN (${\tt q}\geq 2.5$) LAEs      & 602& 1,081 & 1,305 & 87  & 3,075 \\
     \# of ODIN ($2.0 \leq {\tt q} < 2.5$) LAEs & 18& 75    & 65    & 3    & 161   \\
     \# of ODIN (${\tt q}<2.0$)   LAEs           & 58  & 155   & 52    & 4& 269   \\
     \hline     
     \multicolumn{6}{c}{ODIN LAEs with ${\tt q}\geq 2.5$ only } \\
     \hline     
     LAEs                    & 556 & 1,010 & 1,255 & 80 & 2,901 \\
     AGN at target redshift  & 6   & 22    & 18    & 1  & 47    \\
     AGN at lower redshift   & 25  & 28    & 30    & 1  & 84    \\
     Emission line galaxies  & 14  & 21    & 2     & 5  & 42    \\
     star                    & 1   & 0     & 0     & 0  & 1     \\
     \hline
     Sample Purity$^\dagger$ [\%] & 92.2 (82.0--93.1)& 93.3 (77.2--94.5) & 96.2 (88.2--96.4) & 92.0 (85.1--92.6) & -\\
     \hline
    \end{tabular}
    \end{center}
    {\small $^*$ Spectrum quality (${\tt q}$) is defined in Section~\ref{subsec:desi_vi} } \\
    {\small $^\dagger$ Sample purity is defined in Section~\ref{subsec:purity}.}
\caption{A summary of the DESI-ODIN spectroscopic assessments across all fields and redshift windows. The first section denotes the redshift window and provides a general overview of DESI targets. The second section summarizes the number and quality of ODIN LAEs selected. The third section distinguishes the types of objects that were selected as ODIN LAEs that met a quality threshold. Finally, the sample purity shows the face value purity of each sample, as well as the pessimistic and optimistic estimate of the upper and lower bounds of our sample purities.}
\label{tab:summary}
\end{table*}

\begin{figure*}[ht!]
    \centering
    \includegraphics[width=0.8\linewidth,
    ]{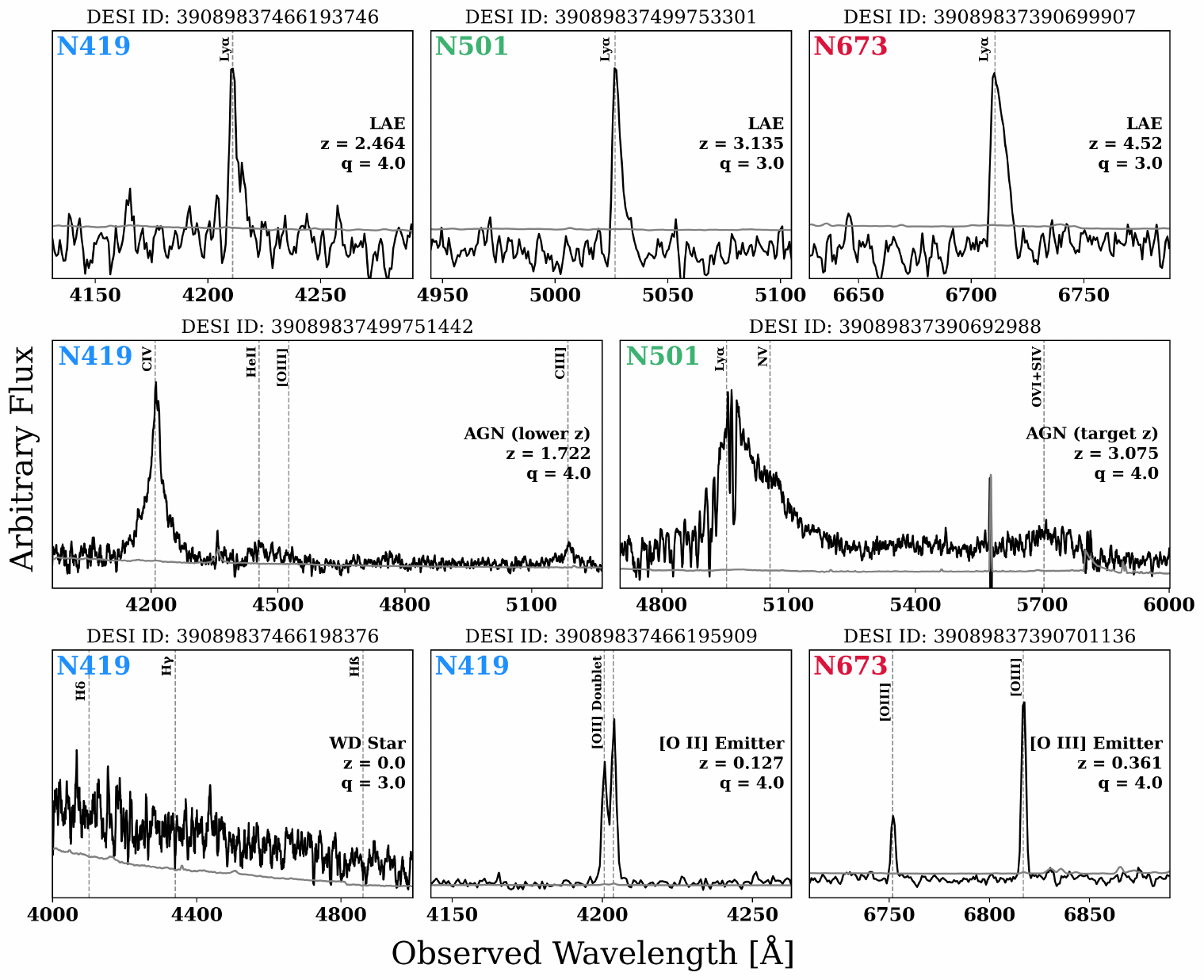}
    \caption{
    Example spectra showcasing the confirmed LAEs and primary contaminants of our sample (AGN at lower-$z$, AGN at target-$z$, Milky Way white dwarfs (WD), [\ion{O}{2}] emitters, and [\ion{O}{3}] emitters). Flux (black) and noise (gray) are boxcar smoothed to exemplify the sources' prominent emission line features, which are denoted by dashed gray lines. The noise spectra were turned into variance spectra and given the same boxcar smoothing. Above each spectrum is the associated DESI target ID, and the panel legend lists the NB filter from which the target was selected, the type of object, redshift ($z$), and quality of the spectrum ({\tt q}). 
    }
    \label{fig:fig2}
\end{figure*}

\begin{table*}[ht!]
  \centering
  \begin{tabular}{l c c c c}
    \hline
     & XMM-LSS $N419$ & COSMOS $N419$ & COSMOS $N501$ & COSMOS $N673$ \\
    \hline\hline
    Tr+ODIN LAEs    & 556  & 1008  & 1255  &  80   \\
    Tr-only LAEs  & 708   & 869   & 670   &  93   \\
    \hline
    Tr-only LAEs -- in Starmasks     & 172 & 331 & 351 & 33 \\
    Tr-only LAEs -- No HSC Broadband & 150 & 88 & 111 & 7 \\
    Tr-only LAEs -- good regions     & 386 & 443 & 200 & 52 \\
    \hline
    Pessimistic completeness\textsuperscript{\(\ddag\)} [\%]   & 58.1 & 68.0 & 84.1 & 58.0 \\
    Optimistic completeness\textsuperscript{\(\ddag\)}  [\%]   & 93.5 & 92.8 & 96.1 & 98.6 \\
    Best-estimate completeness\textsuperscript{\(\ddag\)} [\%] & 92.2 & 91.3 & 95.1 & 98.3 \\
    \hline
  \end{tabular}

  \vspace{2pt}
  {\footnotesize\noindent \(\ddag\)\ Completeness estimates are defined in Section~\ref{sec:completeness}.}
  \caption{The distribution of spec-LAEs across catalogs. The first section compares the number of spec-LAEs selected as LAE candidates by both DESI and ODIN to the number of spec-LAEs selected as LAE candidates by only DESI. The second section shows the spatial diagnostics of spec-LAEs selected by only DESI. The last section displays our estimates of ODIN's completeness.}
\label{tab:spec-lae-distribution}
\end{table*}

\section{Results}\label{sec:results}

In Figure~\ref{fig:fig2}, we compile segments from example DESI spectra of ODIN LAE candidates with robust identification to show the various features that allow the sources to be categorized. The features used to determine spectral types are indicated by vertical dashed lines. Objects identified as LAEs show the characteristic asymmetric line profile \citep[][]{Vitte_2025A&A}, which is caused by the complex radiative transfer of the resonance line. 

The top two rows in Table~\ref{tab:summary} present a summary of our visual inspection campaign. Of the 11,599 sources targeted by DESI, the redshift and spectral types of 7,973 sources are reliably measured. The middle section of the table shows that, of 11,599 sources observed by DESI, only 3,505 (30.2\%) were selected as ODIN LAEs \citep[][see Section~\ref{subsec:comparison} for more detail]{firestone24}. Of those, 3,075 (87.7\%) have measured redshifts and spectral types with quality flag ${\tt q}\geq 2.5$. An additional 161 (4.5\%) have tentative redshift determinations (${\tt q}=2.0-2.5$) and the remaining 269 (7.7\%) have unknown redshifts. Objects with secure quality flags include LAEs and AGN at the target redshift range, AGN and line-emitting galaxies at lower redshift, and a single white dwarf star, as listed in the bottom section of Table~\ref{tab:summary}.  

\begin{figure*}
    \centering
    \includegraphics[width=1\linewidth, alt = {}]{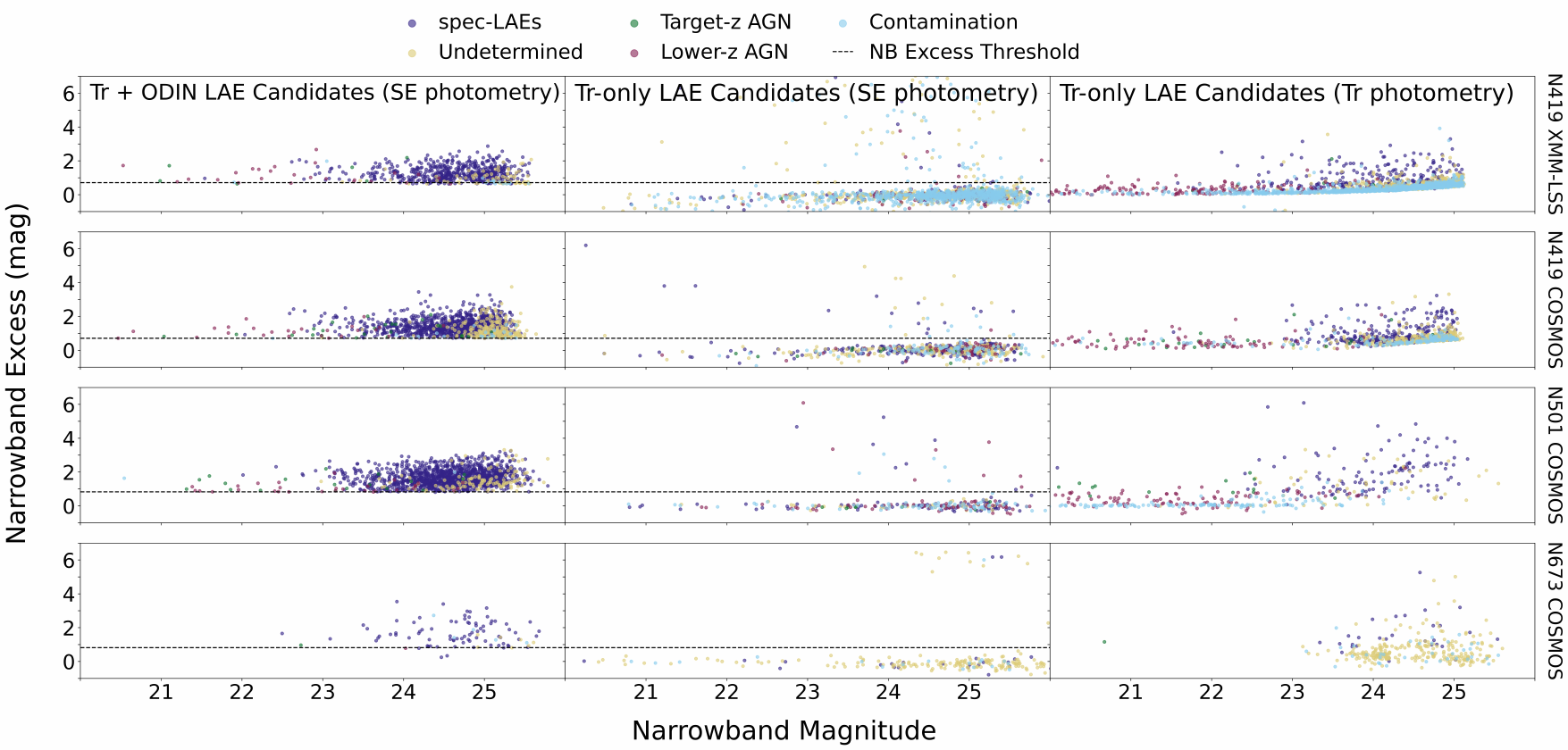}
    \caption{Color-magnitude diagrams for various subsets of objects color-coded by spectral classification determined from the visual inspection campaign (Section~\ref{sec:results}). The left column shows ODIN LAE candidates using SE photometry. The other columns show DESI-targeted sources that are in the ODIN SE catalog but not in the ODIN LAE sample, with the middle column using SE photometry and the right column using Tr photometry. From top to bottom, we display the field/filter combinations: XMM-LSS/$N419$, COSMOS/$N419$, COSMOS/$N501$, and  COSMOS/$N673$. The x-axis shows each object's narrowband (NB) magnitude and the y-axis shows the excess NB flux density in units of magnitudes. Each plot shows confirmed LAEs (red), confirmed AGN in the target redshift range (yellow), confirmed AGN outside the target redshift range (purple), other confirmed emission line galaxies and stars (blue), and targeted objects without identification (orange). The horizontal dashed line in each panel represents a NB excess threshold corresponding to a rest-frame Lyman Alpha equivalent width of 20~\AA.  
    The NB excess in the left and middle columns is calculated with a weighted double broadband magnitude ($rg-N419$, $gr-N501$, or $gi-N673$; see \citealt{firestone24}).  The NB excess for the (rightmost) Tr photometry column is calculated with a single broadband ($g-N419$, $g-N501$, or $r-N673$; described A.~Dey et al. in prep.). We note that the majority of Tr spec-LAEs which were not selected as SE LAE candidates did not have sufficient narrowband excess to satisfy the cut.}
    \label{fig:laecmd}
\end{figure*}

Given the stringent criteria used to select ODIN LAEs, the high ($>$90\%) success rates among spectroscopically-classified sources are not surprising but are nonetheless reassuring. Figure~\ref{fig:laecmd} shows the color-magnitude diagrams for various subsets of sources in the COSMOS ($N419$, $N501$, $N673$) and XMM-LSS ($N419$ only) fields. The left panel shows the Tr+ODIN LAE candidates while the middle and right panels show the Tr-only LAE Candidates utilizing SE photometry and Tr photometry, respectively. The sources are broken up into the following classifications: LAEs in the target redshift range, AGN in the target redshift range, AGN outside the target redshift range, and other contaminants (emission line galaxies and stars). The dashed horizontal lines represent the ODIN LAE color cuts implemented by \citet{firestone24} corresponding to an estimated rest-frame Ly$\alpha$ equivalent width of 20~\AA\null. By construction, all Tr+ODIN LAEs lie above the line, but not all Tr-only LAEs do.

\subsection{Completeness}\label{sec:completeness}

For all samples presented in \autoref{fig:laecmd}, the vast majority of Tr-only LAEs (middle) fall below the narrowband excess cut when plotted with SE photometry. Comparatively, when plotted with Tractor photometry (right), a much higher fraction of Tr-only LAEs satisfy ODIN's rest-frame equivalent width $>20$~{\AA} requirement. For a detailed comparison of the impact of Tractor vs.\ SE photometry in LAE selection, we refer to B.~Thompson et. al. (in prep).  

As noted in ~\autoref{fig:fig1} and detailed in ~\autoref{tab:spec-lae-distribution},  
47.6\%, 70.8\%, and 43.0\% of Tr-only LAEs corresponding to $z$ = 2.45, 3.12, and 4.55 are in regions with no HSC broad-band data or star masks. Of the remaining Tr-only LAEs shown in the middle column of ~\autoref{fig:laecmd}, 75\%, 69\%, and 61\%, respectively, have insufficient narrow-band excess flux density in their SE photometry to be classified as LAEs in the SE catalog. 

Completeness is the measure of how many LAEs ODIN was able to select in a given field with respect to the true total number of LAEs in that field. Since we do not have spectroscopic confirmation of all ODIN targeted objects, we cannot determine the total completeness of the ODIN LAE Sample with certainty. However, we can put upper and lower limits on the completeness of the ODIN LAE sample with three estimation methods. The first divides the number of Tr+ODIN LAEs with spectroscopic confirmation by the total number of spectroscopically confirmed LAEs. This assumes nothing about ODIN LAEs that failed to receive DESI fibers.  It is therefore highly pessimistic, as ODIN found many LAEs that DESI did not target.  

The next method divides the number of ODIN LAEs by the sum of ODIN LAEs and spectroscopically-confirmed Tr-only LAEs in good data regions. This assumes that all ODIN LAEs without DESI targeting are bona fide LAEs, which makes it optimistic. The last method \textbf{allows us to make our} best estimate for completeness by dividing the number of ODIN LAEs multiplied by the sample purity by the sum of ODIN LAEs multiplied by sample purity plus all spectroscopically-confirmed Tr-only LAEs in good data regions. This corrects the untargeted portion of the ODIN LAE catalog by our measured purity fraction, which is the fairest estimate we can offer. Due to the high ($>$90\%) purity, the optimistic and best estimates for completeness are quite close, as can be seen in ~\autoref{tab:spec-lae-distribution}.  

\subsection{Sample Purity}\label{subsec:purity}
The DESI-ODIN spectroscopy program is primarily motivated to understand the purity of the sample of LAE candidates and the demographics of various interloper populations. Such information is crucial to statistical analyses (e.g., clustering, luminosity function) as it provides the necessary corrections for these measurements. We estimate sample purity in two different ways. First, if we assume that our spectroscopic failures are due to the faintness of the sources and the demographics of our targets do not appreciably change as a function of brightness, our sample purity is simply the number of confirmed ODIN LAEs divided by the number of DESI targets with ${\tt q} \geq 2.5$, as given in Table~\ref{tab:summary}. This yields a purity of 93.0\%, 96.2\%, and 92.0\% for the $N419$-, $N501$-, and $N673$-selected LAEs, respectively. 

However, this assumption is unlikely to be correct. As discussed in Section~\ref{subsec:demographics}, the data strongly suggest that the relative budget of interlopers changes as a function of optical brightness, and contamination by continuum-only galaxies likely increases at dim narrow-band magnitudes. To mitigate this uncertainty, we consider the worst-case scenario: i.e., if all targeted sources that did not yield reliable redshifts are low-redshift contaminants, we can estimate the lower limit by dividing the number of confirmed LAEs by the number of ODIN LAEs targeted by DESI. Conversely, the most optimistic estimate can be made by assuming that all unidentified sources are LAEs. The ranges of purity estimated this way are listed in the bottom row of Table~\ref{tab:summary}. 

In reality, the true contamination rate lies somewhere between the considered possibilities. Given the relatively short exposure times used for these observations, some sources evading identification are likely real LAEs. Additionally, it is worth noting that about a quarter of all known contaminants are broad-lined AGN at a similar redshift range to our LAEs and thus have similar colors. All in all, the visual inspection results suggest that the integrated purity of the sample of ODIN LAEs is quite high, with a well-understood contamination rate. 

As noted in Section~\ref{subsec:desi_vi}, the effective DESI exposure times were not uniform across the ODIN targets, with a small subset of sources observed for less than 1.3 hr. Because shorter integrations could reduce the spectroscopic success rate, we repeated the purity calculation excluding these shallow observations. The resulting purities for the \textit{N419}, \textit{N501}, and \textit{N673} samples (92.5\%, 96.1\%, and 90.9\%, respectively) are statistically indistinguishable from those derived using the full sample. We therefore conclude that variations in exposure time do not significantly bias our inferred contamination rates and adopt the full spectroscopic catalog in the subsequent analysis.

\begin{figure*}
    \centering
    \includegraphics[width=0.8\linewidth, alt = {}]{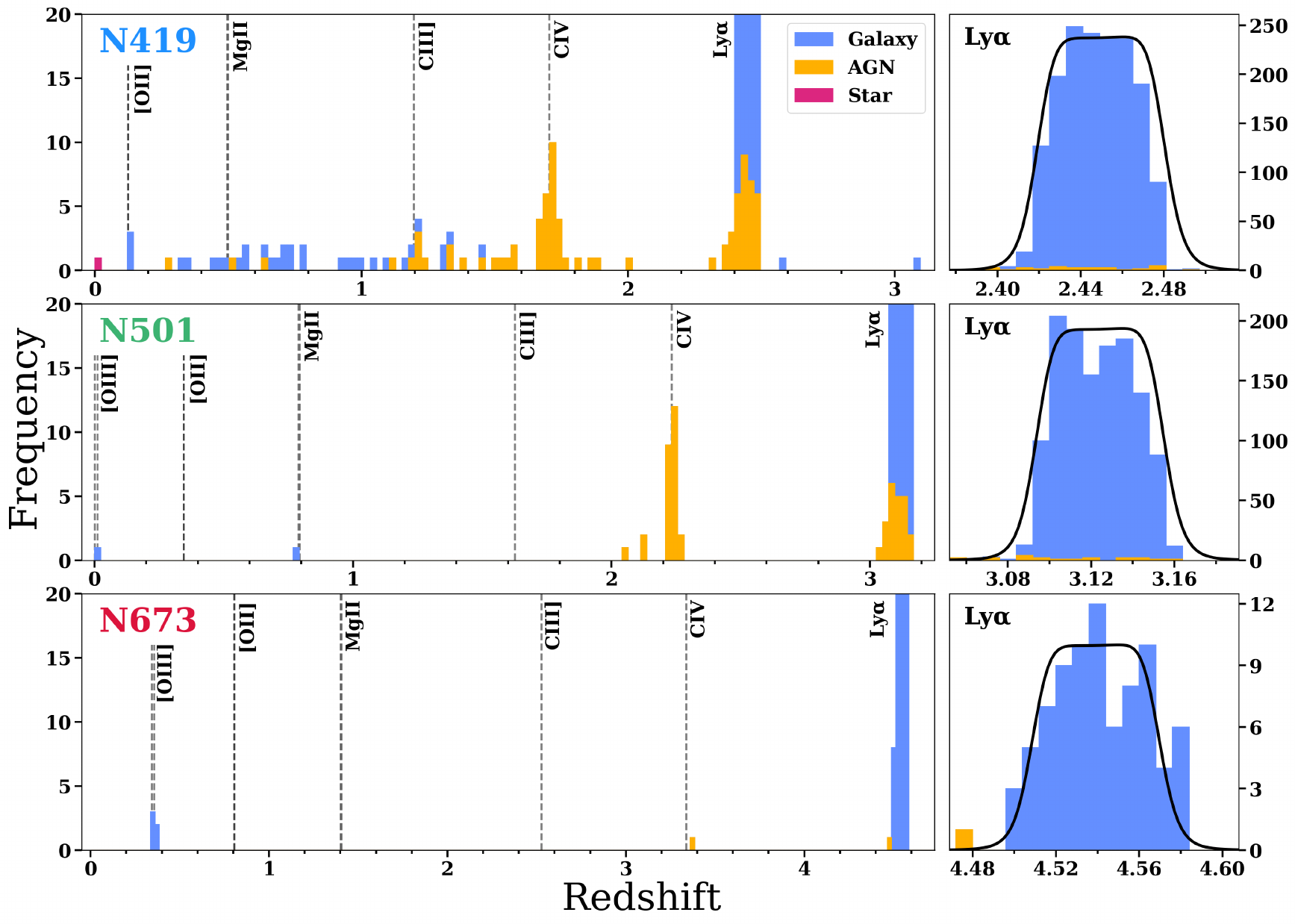}
    \caption{Redshift distribution of DESI-confirmed (${\tt q}\geq 2.5$) ODIN sources observed across all three NB filters. \textit{Left:} A histogram of the entire redshift range showing contaminants that are not within the Ly$\alpha$ redshift window defined by the transmission curve of the NB filter. The dashed lines indicate the redshifts associated with common non-Ly$\alpha$ emission lines. \textit{Right:} Zoomed-in redshift histogram of Ly$\alpha$ emitting objects. The black curves represent the transmission functions of the NB filters.}
    \label{fig:z_hist}
\end{figure*}

\subsection{The Demographics of ODIN LAE candidates}\label{subsec:demographics}
In Figure~\ref{fig:z_hist}, we present the redshift histograms of ODIN LAE candidates. The left panels display the full redshift range, while the right panels focus on the Ly$\alpha$ redshift range anticipated for the three ODIN filters \citep[see][]{lee24}: $N419$ (top), $N501$ (middle), and $N673$ (bottom). For the confirmed LAEs, their redshift distribution is in reasonable agreement with the transmission curves expected by the filters, as shown in the figure. As mentioned previously,  LAEs make up $\approx$93.0\% ($N419$), 96.2\% ($N501$), and 92.0\% ($N673$) of the confirmed sources in each sample, with broad-line AGN at the same redshift range adding another 1--2\%.

For the $N501$- and $N673$-selected LAEs, the nature of interloper populations is straightforward: they are [O~{\sc iii}], Mg~{\sc ii}, and C~{\sc iv} emitters, as indicated by the vertical dashed lines in the figure. The relatively broad spread in redshifts around these lines is because they are typically QSOs showing broad line widths. For [O~{\sc iii}] emitters, the redshift range is slightly broad because either or both of the doublet (4959~\AA\ and 5007~\AA) can fall into a given NB filter, masquerading as an LAE. It is worth noting that the contamination by [O~{\sc ii}] emitters is absent in both $N501$ and $N673$ samples even though they are often considered as primary contaminants \citep[][]{Leung_2017ApJ}. 

In contrast to the other two, the redshift values of the $N419$ LAE interlopers include but are not limited to those belonging to several strong lines ([O~{\sc ii}], Mg~{\sc ii}, C~{\sc iii}], and C~{\sc iv}); they appear to form a continuous range between $z=0.2-1.5$. Of the 30 interlopers (17 and 13 in COSMOS and XMM-LSS, respectively; see Table~\ref{tab:summary}) classified as spectral type {\tt  Galaxy}, 17 are relatively continuum-faint star-forming galaxies at $z=0.2-0.8$ identified via [O~{\sc ii}] emission. For reference, [O~{\sc ii}] falls into the $N419$ filter at $z=0.125$. At $z=0.5-0.8$ and $0.15-0.40$, the [O~{\sc ii}] and [O~{\sc iii}] doublet, respectively, will fall into the $r$ band. In the adopted double broad-band scheme used for the ODIN LAE selection \citep[see more detail in][]{firestone24}, the continuum level would be underpredicted for [O~{\sc ii}] emitters in those redshift ranges, allowing some of them to scatter into the LAE sample.

The remaining 13 galaxies at $z = 1.0 - 1.5$ are more difficult to understand. Their spectra do not show any features in the $N419$ wavelength range. Instead, their redshifts are mostly determined by the detection of the resolved [O~{\sc ii}] doublet at $\lambda_{\rm obs} = 7450\text{--}9200$~\AA. Because these lines can contaminate the $i$- and  $z$ bands, it would not affect our LAE selection, which uses $N419$, $g$, and $r$ only. The only line that could plausibly contaminate the $r$-band flux is Mg~{\sc ii}, but it is not detected in the data. We speculate that these 13 galaxies may obey an extinction law with a strong UV 2175~\AA\ bump, which depresses the $g$-band flux, causing the continuum level to be underestimated. While existing studies suggest that dust extinction curves change as a function of spectral types where the 2175~\AA\ bump becomes more prominent in more evolved galaxies \citep[see, e.g.,][]{conroy10,scoville15,zeimann15}, our data already suggest that they constitute less than 0.6\% of the entire $N419$-selected sample. If our hypothesis is correct, the implication would be that more inclusive color cuts and/or single-BB-based LAE selection methods 
could lead to a higher contamination rate by these rare galaxies in LAE samples.

\begin{figure*}[t!]
    \centering
    \includegraphics[width=0.8\linewidth, alt = {}]{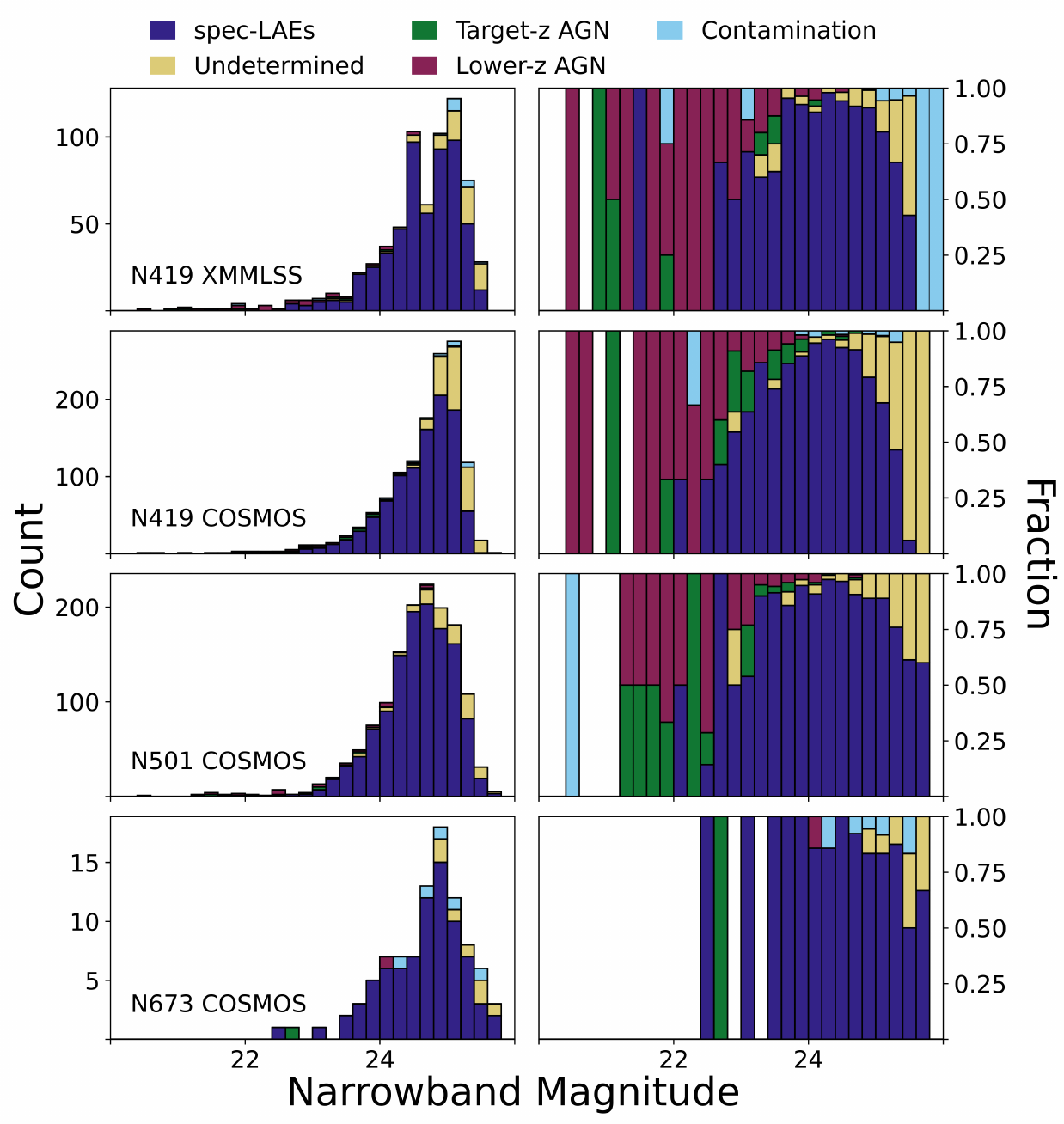}
    \caption{Histograms displaying Tr+ODIN LAE candidates sorted by narrowband magnitude and colored by classification. Narrowband magnitude is displayed in bins of size 0.2 and ranges from 20 to 26.2. The colors are consistent with those in Figure 3. The left column displays the raw histograms stacked on top of one another. The right column displays the normalized histograms for the same bin sizes. At dimmer magnitudes, a higher percentage of the ODIN LAE Candidates are targeted objects without identification, and at brighter magnitudes, there is a higher percentage of contaminants.}
    \label{fig:interloper_fraction}
\end{figure*}

The availability of spectral types and redshifts allows us to look at more detailed demographics of interloper populations at different brightness levels. In Figure~\ref{fig:interloper_fraction}, we display the total and relative contribution of each interloper type as a function of NB magnitudes, shown separately for each filter/field combination. As before, we distinguish AGN at the ODIN redshift range from those at lower redshifts in this figure. 

At the bright end (${\rm NB} \lesssim 22.5$), AGN generally dominate, although their total number is negligible compared to that of confirmed LAEs. LAEs take up at least half of the number in each magnitude bin at ${\rm NB}\approx 23$ and reach above 90\% at its peak around ${\rm NB}\approx 24$. Even with relatively short exposures taken by DESI, its efficiency is high with a success rate around 50\% at NB magnitude $\approx$25.5. The fractional contribution of low-$z$ star-forming galaxies is very small for $N501$ LAEs with a single [O~{\sc iii}] and Mg~{\sc ii} emitter each. However, it is higher for $N419$ and $N673$ LAEs. As discussed previously, for $N419$, this appears to be connected to strong oxygen emission lines falling in and contaminating the $r$-band flux densities, affecting our ability to perform accurate continuum subtraction at the Ly$\alpha$ wavelength. 

\subsection{Color-color diagrams}

\begin{figure*}[ht!]
    \centering
    \includegraphics[width=0.9\linewidth, alt = {}]{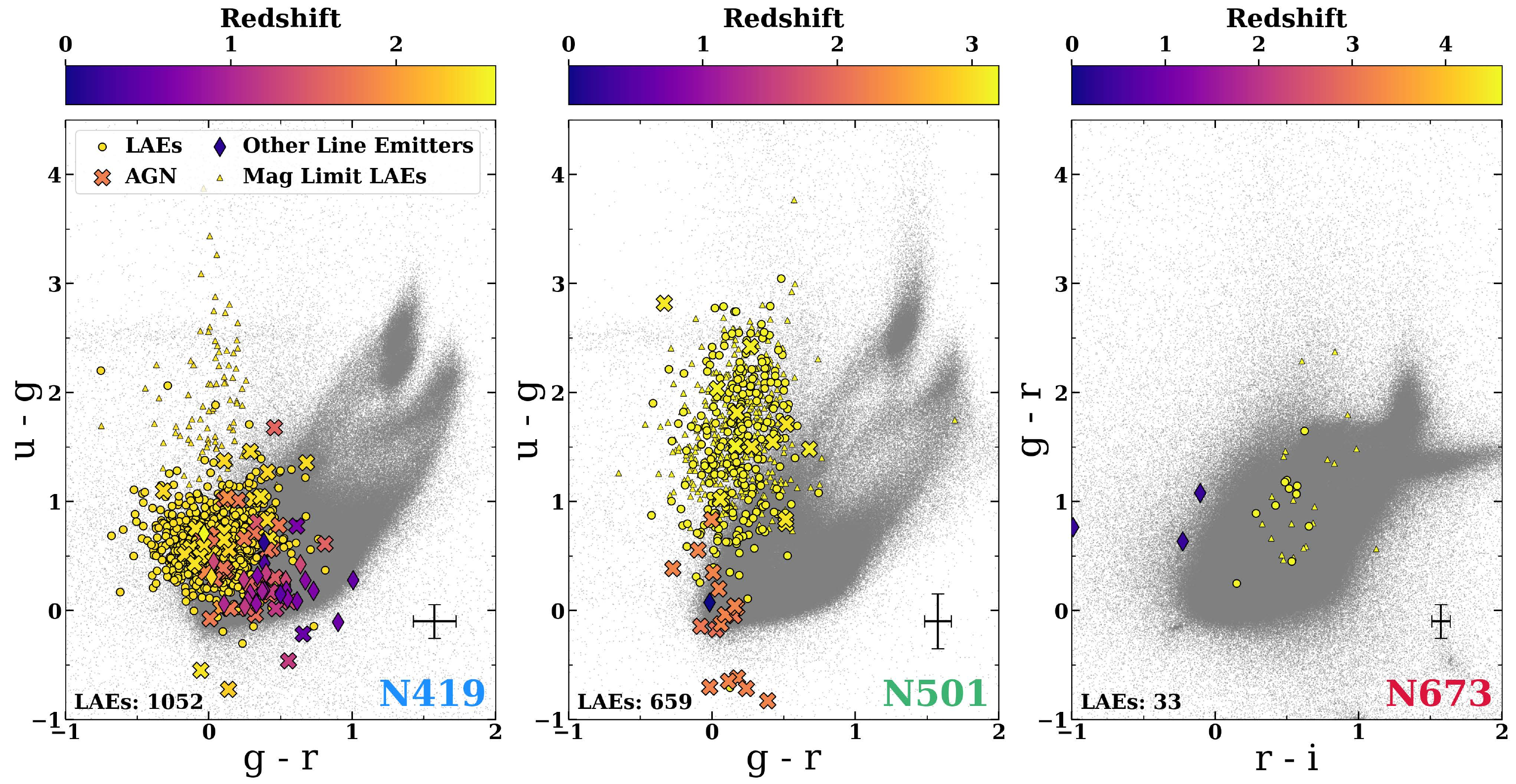}
    \caption{Color-color diagrams of DESI-confirmed ODIN sources with sufficient broad-band measurements. These sources are color-coded by redshift according to the colorbar above each subplot. The gray points are all ODIN sources within the SE photometric catalogs. The small yellow triangles are the upper color limits of ODIN LAEs with sufficient data in two redder bands (g and r for \textit{N419}/\textit{N501}, r and i for \textit{N673}), but insufficient measurements in the bluest bands (u for \textit{N419}/\textit{N501}, g for \textit{N673}). In the bottom right corners are example LAEs with the median x- and y-axis error bars.
    }
    \label{fig:two_color_diagram}
\end{figure*}

In Figure~\ref{fig:two_color_diagram}, we show the two-color diagrams where the color shown in the ordinate straddles the Lyman break. As a subset of star-forming galaxies, LAEs are expected to occupy a region separated from the main locus of galaxies when the colors can be measured robustly. We require that the filters used to compute the UV continuum color ($g-r$ for $N419$ and $N501$ LAEs, and $r-i$ for $N673$ LAEs) have a minimum S/N of 3. For the filter that samples the Lyman limit, we replace the measured flux with the 2$\sigma$ limiting flux\footnote{The $u$ band data is from the CLAUDS survey \citep{sawicki19}, which covers a subsection of each of the HSC SSP deep fields. The $2\sigma$ limiting magnitude measured within 2" circular apertures is 27.7~AB.} whenever S/N is equal to or less than 2. The number of LAEs that are shown in Figure~\ref{fig:two_color_diagram} is shown in the bottom left corner of each panel (66\% and 53\% of the confirmed $N419$ and $N501$ LAEs, respectively). 

The locations occupied by the confirmed LAEs are reassuringly consistent with the measured colors of Lyman break galaxies at similar redshift \citep[e.g.,][]{giavalisco02,steidel03,steidel04}. At $z=2.4$, the Lyman limit falls only partially into the $u$ band, producing smaller flux decrements than those expected at $z=3.1$. Similarly, at $z=4.5$, the Lyman limit redshifts to 5016~\AA\ and the $g$-band transmission cut off steeply at $\approx$5500~\AA. Thus, the $g-r$ color is moderately reddened. Additionally, in all cases, the $g$- and $r$-band fluxes include Ly$\alpha$ emission at observed equivalent widths of $\gtrsim 70$~\AA, 80~\AA, and 110~\AA\ for $N419$, $N501$, and $N673$ LAEs, respectively.

As expected, the majority of lower-redshift interlopers can be distinguished by these two-color diagrams as they show much bluer colors in the ordinate. At $z=3.1$, the main contaminants C~{\sc iv}-emitting QSOs show very blue $u-g$ colors. Soon, Rubin's Legacy Survey for Space and Time \citep[LSST; e.g.][]{ivezic2019ApJ} will obtain deep $ugrizy$ imaging of all ODIN fields, making it possible to eliminate most of the low-z interlopers based on broad-band colors alone. 

\section{Conclusions} \label{sec:conclusions}
In this paper, we presented DESI observations of Ly$\alpha$-emitting galaxy candidates identified by the ODIN survey at $z = 2.45 \pm 0.03$, $3.12 \pm 0.03$, and $4.55 \pm 0.04$ \citep[][]{lee24}. These observations targeted a total of 11,599 LAE candidates in the extended COSMOS and XMM-LSS field. A visual inspection campaign resulted in a robust determination of spectral type and redshift for  7,973 objects. Using these data, we validated the ODIN LAE selection (Section~\ref{subsec:desi-odin-lae}) and obtained an accurate picture of what types of sources contaminate our samples. Our main findings are summarized below. 

\begin{enumerate}

\item The ODIN selection criteria described in \citet{firestone24} result in highly pure LAE samples at all three redshifts (Table~\ref{tab:summary}). Of all sources with robust identification, 92.8\%, 96.2\%, and 92.0\% are confirmed in LAE at $z\approx 2.45$, $3.12$, and $4.55$, respectively. An additional 1--2\% are broad-line AGN at a similar redshift. Even in the most pessimistic scenario (i.e., if all sources with failed redshift identification are contaminants, which is unlikely), the contamination rate is less than 20\%, 12\%, and 15\% at these redshift ranges. The redshift distribution of confirmed LAEs is broadly consistent with that expected from the ODIN NB transmission curve.

\item Interlopers in our LAE samples include broad-line AGN at redshifts similar to our LAEs found via their Ly$\alpha$, broad- and narrow-line AGN at lower redshift via C~{\sc iii}], C~{\sc iv}, and Mg~{\sc ii}, star-forming galaxies with strong oxygen lines ([O~{\sc ii}] and [O~{\sc iii}]), and a single white dwarf star (Figure~\ref{fig:z_hist}).  All of the $N501$ and $N673$ interlopers have redshift ranges consistent with one of these lines falling into these filters. 

\item The $N419$ sample interlopers show a more complex behavior, with an additional contribution ($\sim 2$\%) from star-forming galaxies that form a semi-continuous redshift distribution at $z=0.2-0.8$ and $z=1.0-1.5$. DESI spectra of these sources show no obvious spectral feature within the $N419$ wavelength range. For the former group, [O~{\sc ii}] or [O~{\sc iii}] fall into the $r$ band, likely resulting in the underestimation of the continuum flux at Ly$\alpha$ wavelength when the double broad-band continuum estimation is employed, with a resulting overestimation of the line equivalent width. We speculate that the $1<z<1.5$ set may represent a rare class of low-mass star-forming galaxies that possess dust with an unusually strong 2175~\AA\ extinction bump, causing the continuum level to be underestimated in the $g$-band filter.  

\item The sheer size of the DESI-ODIN spectroscopic sample allows us to robustly determine the relative budget of contaminants as a function of NB magnitudes (Figure~\ref{fig:interloper_fraction}). Such information is essential in statistical analyses such as luminosity functions, clustering, and protocluster detections by providing an appropriate correction factor. Finally, we show that a great majority of these interlopers will be easily rejected based on existing deep $ugri$ coverage. This bodes well for the future ODIN LAE selection as most ODIN fields are LSST Deep Drilling Fields that will receive 10-year LSST imaging depth within the first year of Rubin operations.

\item A further question that we have been able to address is that of the completeness of the ODIN LAE samples. The more inclusive selection of the Tr LAE samples means that a significant fraction of the spec-LAEs represent objects beyond the ODIN LAE samples. However, the vast majority of spec-LAEs beyond the ODIN LAE sample were missed through conservative star masking or a lack of HSC Broadband data, i.e., they appear in regions where ODIN did not try to find LAEs.  Our best estimate for completeness of the ODIN LAE samples is  91-92\% at $z=2.45$, 95\% at $z=3.12$, and (although the highest redshift offers limited statistics) 98\% at $z=4.55$.  
\end{enumerate}

\section*{Data Availability}
All data shown in figures are available on Zenodo (doi: \href{https://zenodo.org/records/18460301}{https://doi.org/10.5281/zenodo.18460301}).

\section*{Acknowledgments}
We acknowledge financial support from the National Science Foundation (NSF) under Grant Nos. AST-2206705, AST-2408359, and AST-2206222, and from the Ross-Lynn Purdue Research Foundations. EG and NF acknowledge support from DOE grant DE-SC0010008.  Ho Seong Hwang acknowledges the support of the National Research Foundation of Korea (NRF) grant funded by the Korean government (MSIT), NRF-2021R1A2C1094577, and Hyunsong Educational \& Cultural Foundation.

This material is based upon work supported by the U.S. Department of Energy (DOE), Office of Science, Office of High-Energy Physics, under Contract No. DE–AC02–05CH11231, and by the National Energy Research Scientific Computing Center, a DOE Office of Science User Facility under the same contract. Additional support for DESI was provided by the U.S. National Science Foundation, Division of Astronomical Sciences under Contract No. AST-0950945 to the NSF's National Optical-Infrared Astronomy Research Laboratory; the Science and Technology Facilities Council of the United Kingdom; the Gordon and Betty Moore Foundation; the Heising-Simons Foundation; the French Alternative Energies and Atomic Energy Commission (CEA); the National Council of Humanities, Science and Technology of Mexico (CONAHCYT); the Ministry of Science, Innovation and Universities of Spain (MICIU/AEI/10.13039/501100011033), and by the DESI Member Institutions: \url{https://www.desi.lbl.gov/collaborating-institutions}. Any opinions, findings, and conclusions or recommendations expressed in this material are those of the author(s) and do not necessarily reflect the views of the U. S. National Science Foundation, the U. S. Department of Energy, or any of the listed funding agencies.

The authors are honored to be permitted to conduct scientific research on I'oligam Du'ag (Kitt Peak), a mountain with particular significance to the Tohono O'odham Nation.

These data were obtained and processed as part of the CFHT Large Area U-band Deep Survey (CLAUDS), which is a collaboration between astronomers from Canada, France, and China described in Sawicki et al. (2019, [MNRAS 489, 5202]).  CLAUDS data products can be accessed from https://www.clauds.net. CLAUDS is based on observations obtained with MegaPrime/MegaCam, a joint project of CFHT and CEA/DAPNIA, at the CFHT, which is operated by the National Research Council (NRC) of Canada, the Institut National des Science de l'Univers of the Centre National de la Recherche Scientifique (CNRS) of France, and the University of Hawaii. CLAUDS uses data obtained in part through the Telescope Access Program (TAP), which has been funded by the National Astronomical Observatories, Chinese Academy of Sciences, and the Special Fund for Astronomy from the Ministry of Finance of China. CLAUDS uses data products from TERAPIX and the Canadian Astronomy Data Centre (CADC) and was carried out using resources from Compute Canada and Canadian Advanced Network For Astrophysical Research (CANFAR). 

\bibliography{myrefs}{}
\bibliographystyle{aasjournal}



\end{document}